\documentclass[aps,prl,twocolumn,showpacs,groupedaddress]{revtex4}

\usepackage{graphicx}
\usepackage{amssymb}

\begin{document}

\preprint{}

\title{Dominant role of oxygen vacancies in electrical properties of unannealed LaAlO$_{3}$/SrTiO$_{3}$ interfaces}

\author{Z. Q. Liu$^{1,2}$}

\author{L. Sun$^{1}$}

\author{Z. Huang$^{1}$}

\author{C. J. Li$^{1,3}$}

\author{S. W. Zeng$^{1,2}$}

\author{K. Han$^{1,2}$}

\author{W. M. L\"{u}$^{1}$}

\author{T. Venkatesan$^{1,2,3,4}$}

\author{Ariando$^{1,2}$}

\altaffiliation[Email: ]{ariando@nus.edu.sg}

\affiliation{$^1$NUSNNI-Nanocore, National University of Singapore, 117411 Singapore}

\affiliation{$^2$Department of Physics, National University of Singapore, 117542 Singapore}

\affiliation{$^3$National University of Singapore (NUS) Graduate School for Integrative Sciences and Engineering, 28 Medical Drive, 117456 Singapore}

\affiliation{$^4$Department of Electrical and Computer Engineering, National University of Singapore, 117576 Singapore}

\date{\today}

\begin{abstract}
We report that in unannealed LaAlO$_{3}$/SrTiO$_{3}$ heterostructures the critical thickness for the appearance of the two-dimensional electron gas can be less than 4 unit cell (uc), the interface is conducting even for SrTiO$_{3}$ substrates with mixed terminations and the low-temperature resistance upturn in LaAlO$_{3}$/SrTiO$_{3}$ heterostructures with thick LaAlO$_{3}$ layers strongly depends on laser fluence. Our experimental results provide fundamental insights into the different roles played by oxygen vacancies and polarization catastrophe in the two-dimensional electron gas in crystalline LaAlO$_{3}$/SrTiO$_{3}$ heterostructures.
\end{abstract}



\maketitle


The two-dimensional electron gas (2DEG) at the LaAlO$_{3}$/SrTiO$_{3}$ (LAO/STO) interface has been intensively investigated since its discovery [1]. Whether the origin of the 2DEG is polarization catastrophe [2] or oxygen vacancies has been hotly debated. We have recently clarified the controversy by a series of experiments comparing the electrical properties of amorphous and crystalline LAO/STO heterostructures [3]. It was found that the 2DEG can be created via either oxygen vacancies (2DEG-V) when the layer on top of STO contains elements with strong oxygen affinity such as Al [4], or polarization catastrophe (2DEG-P). In amorphous LAO/STO heterostructures, the 2DEG-V is the only contributor to interface conduction; in crystalline LAO/STO heterostructures that have not been oxygen-annealed in a high oxygen pressure (typically 0.4 to 1 bar) after deposition, both the 2DEG-V and the 2DEG-P contribute to the interface conduction; nevertheless, the 2DEG-P alone accounts for the interface conductivity in oxygen-annealed crystalline LAO/STO heterostructures. It was also found that a high degree of crystallinity of the LAO layer is essential for the 2DEG-P formation in crystalline LAO/STO heterostructures [3]. More importantly, the 2DEG-V is thermally activated as the donor level of oxygen vacancies in STO single crystals is $\sim$4 meV below the STO conduction band and shows the carrier freeze-out effect [3,5], while the 2DEG-P is degenerate.

The critical thickness for the appearance of interface conductivity in oxygen-annealed crystalline LAO/STO interfaces is 4 uc [6], which clearly supports the polarization catastrophe model over oxygen vacancies [7]. It is then intriguing to explore whether the critical thickness of 4 uc is still valid for unannealed crystalline LAO/STO interfaces. To do so, we fabricated crystalline LAO/STO heterostructures with LAO layers 3-uc-thick. Pulsed laser deposition (KrF laser \emph{$\lambda$} = 248 nm) equipped with reflection high energy electron diffraction (RHEED) was utilized. A single-crystal LAO target was utilized and the target-substrate distance was 55 mm. During depositions of 3-uc-thick LAO films on TiO$_{2}$-terminated STO substrates at 750 $^{\circ}$C, the oxygen pressure was kept at 10$^{-4}$ Torr. The laser fluence was fixed at 1.3 J/cm$^{2}$. At this oxygen pressure we do not introduce the three-dimensional electrons in STO substrates [8]. After deposition, samples were cooled down to room temperature in the deposition oxygen pressure. Electrical properties of such heterostructures were examined by a Quantum Design physical property measurement system and electrical contacts on 5$\times$5 mm$^{2}$ samples were made by wire bonding via Al wires. Sheet resistance and Hall measurements were performed in the Van der Pauw geometry.

To achieve a smaller critical thickness, we fabricated 3-uc LAO films on TiO$_{2}$-terminated STO substrates in a relatively low oxygen partial pressure of 10$^{-4}$ Torr at 750 $^{\circ}$C. TiO$_{2}$-terminated STO substrates were obtained by buffered-HF acid treatment followed by an annealing process in the 1 bar oxygen flow at 950 $^{\circ}$C for 2 h. Such as-deposited 3-uc LAO/STO heterostructures were determined to be electrically measurable with room-temperature sheet resistance of the order of 10$^{5}$ $\Omega$/$\square$  and metallic at low temperatures [Fig. 1(a)]. This conductivity arises from the 2DEG-V as the 2DEG-P requires a minimum LAO thickness of 4 uc [3,6]. This was further confirmed by \emph{ex situ} oxygen annealing at 600 $^{\circ}$C in 1 bar of oxygen flow for 1 h, after which the interface conduction completely disappeared [Fig. 1(b)].  In contrast, LAO/STO heterostructures with the LAO layer thicker than 3 uc remain conducting after oxygen annealing.


\begin{figure}
\includegraphics[width=3.3in]{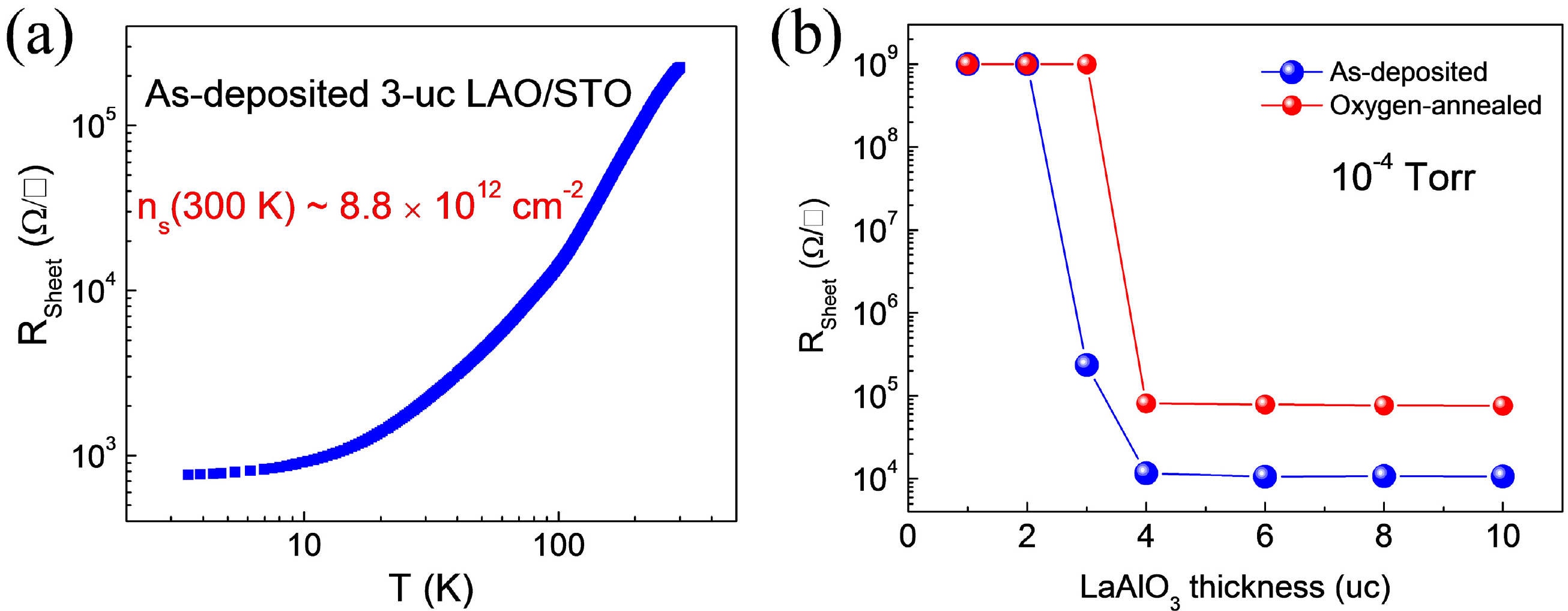}
\caption{\label{fig1}(Color online) (a) Temperature-dependent sheet resistance of an as-deposited 3-uc LAO/STO heterostructure deposited in 10$^{-4}$ Torr oxygen partial pressure and 750 $^{\circ}$C. (b) Room-temperature sheet resistance of LAO/STO heterostructures deposited at 10$^{-4}$ Torr as a function of the LAO layer thickness before and after oxygen annealing.}
\end{figure}

The room-temperature sheet carrier density of an as-deposited 3-uc LAO/STO heterostructure is 8.8$\times$10$^{12}$ cm$^{-2}$, which is one order of magnitude smaller than that of as-deposited 10-uc LAO/STO heterostructures [3]. The 2DEG-V formation is due to the interface chemical reaction [3,4,9], where Al atoms and ions in the pulsed laser deposition plume are chemically active to react with oxygen from both atmosphere and STO surface. There is also a critical thickness of LAO films for the 2DEG-V formation, which is needed for percolation of electrons (Mott carrier limit) originating from oxygen vacancies. However, it is not universal and strongly depends on oxygen pressure, laser fluence and substrate-target distance. In our experimental setup, the critical thickness for the 2DEG-V formation in 10$^{-4}$ Torr oxygen pressure was found to be 1 nm [3]. The appearance of the 2DEG-V in as-deposited 3-uc LAO/STO heterostructures is therefore consistent with our previous results and in good agreement with the previous study by Cancellieri \emph{et al.} [10]. More importantly, it clearly demonstrates that the 2DEG-V formation in unannealed crystalline LAO/STO heterostructures does not need 4-uc LAO and is independent of the polarization catastrophe.

The 2DEG-P formation in crystalline LAO/STO heterostructures is known to be sensitive to the STO surface termination [1]. An \emph{n}-type LaO/TiO$_{2}$ interface is conducting while a \emph{p}-type AlO$_{2}$/SrO interface has much higher resistance [11]. It is then also intriguing to know what if a 10-uc LAO layer is deposited on a randomly-terminated STO substrate. It seems obvious that the conductivity depends on the area ratio between TiO$_{2}$-terminated and SrO-terminated regions. However, from our atomic force microscopy (AFM) studies, nearly all as-received (001)-oriented STO substrates have mixed terminations with half unit-cell steps and disordered step edges[Figs. 2(a)-(b)], where a percolative connection of any termination (either TiO$_{2}$ or SrO) over a macroscopic scale is lacking. This is consistent with the AFM studies of as-received STO substrates by Huijben [12].

As a large number of groups fabricate LAO/STO heterostructures at relatively high oxygen pressure such as 10$^{-3}$ Torr (this was expected to minimize the effect of oxygen vacancies, but it turned out to be not ture [3]), we fabricated LAO/STO heterostructures with LAO layer thickness of $\sim$10 uc on randomly-terminated STO substrates in 10$^{-3}$ Torr oxygen pressure at 750 $^{\circ}$C. The deposition rate of the LAO layer was calibrated by RHEED oscillations of LAO films grown on TiO$_{2}$-terminated STO substrates. During deposition, the laser fluence was kept at 1.3 J/cm$^{2}$. After deposition, samples were cooled down to room temperature in the deposition pressure.

\begin{figure}
\includegraphics[width=3.3in]{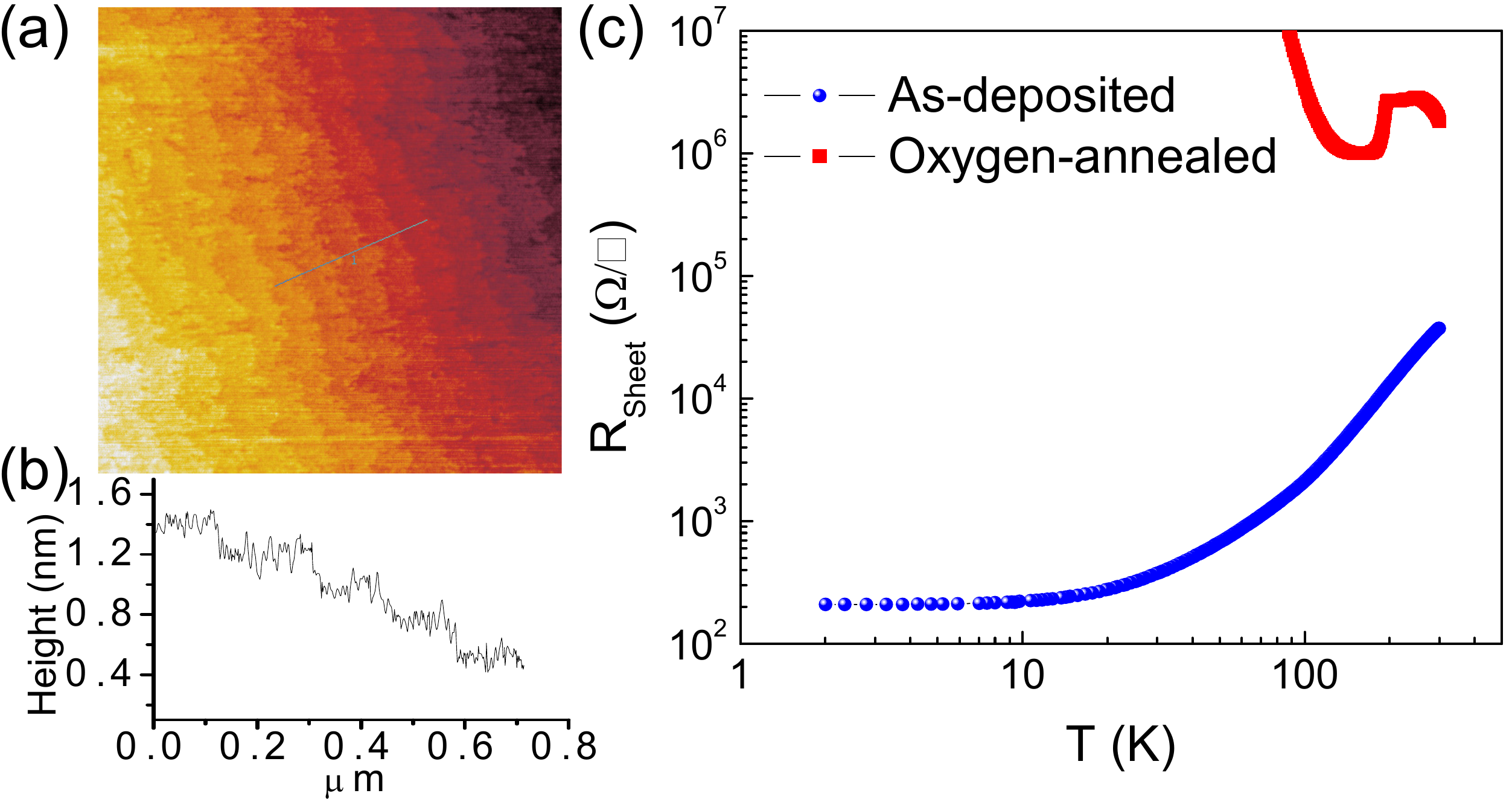}
\caption{\label{fig2} (Color online) (a) AFM image of an as-received STO substrate. (b) Height profile of the line marked by ``1'' in the (a). (c) Temperature-dependent sheet resistance of a 10-uc LAO/STO heterostructure fabricated on a STO substrate with mixed terminations before and after oxygen annealing.}
\end{figure}

It was found that such as-deposited LAO/STO heterostructures are conductive as well [Fig. 2(c)]. After oxygen annealing in a 1 bar oxygen gas flow at 600 $^{\circ}$C for 1 h, the sheet resistance of such LAO/STO interfaces increases by a factor of $\sim$10$^{2}$. Consequently, the temperature-dependent transport behavior resembles that of a \emph{p}-type LaO/SrO interface [11]. As-deposited heterostructures fabricated by depositing 10-uc LAO on SrO-terminated STO at 2$\times$10$^{-5}$ Torr oxygen pressure [12] however show insulating characteristic [11,12], which implies that the SrO termination does not support a 2DEG of any kind. So in the case of mixed terminations, the surprising result that one sees metallic behavior in as-deposited heterostructures indicates the 2DEG-V formation and also suggests that the oxygen vacancies created in the SrO termination may play some role in increasing the percolative path of TiO$_{2}$. However, after annealing, the same admixture of TiO$_{2}$ and SrO regions is unable to support a conducting path for the 2DEG-P. In contrast, an oxygen-annealed 10-uc LAO/STO heterostructure fabricated on a TiO$_{2}$-terminated STO substrate is metallic [3], characteristic of the \emph{n}-type 2DEG-P.

Another striking feature in the crystalline LAO/STO interface system is the low-temperature resistance minimum appearing in LAO/STO heterostructures with thick LAO layers deposited at relatively high oxygen pressure such as 10$^{-3}$ Torr [13]. Whether the low-temperature insulating-like behavior is due to the Kondo effect [13] or weak localization [14] is still controversial. We fabricated a series of 20-uc LAO/STO heterostructures based on TiO$_{2}$-terminated STO substrates in 10$^{-3}$ Torr oxygen pressure with different laser fluence. After deposition, samples were cooled down to room temperature in the deposition oxygen pressure.

20-uc LAO/STO heterostructures deposited with the laser fluence of 1.3 J/cm$^{2}$ exhibit atomically flat surface with the STO substrate steps maintained during the layer-by-layer growth of the LAO layer [Fig. 3(a)]. No crack in LAO films was seen both in optical microscopy and AFM measurements. The electrical properties of 20-uc LAO/STO heterostructures were examined by sheet resistance measurements. It was found that all the heterostructures exhibit metallic behavior at high temperatures. However, samples fabricated with lower laser fluence possess a resistance upturn at low temperatures while samples deposited with higher laser fluence are metallic over the entire temperature range [Fig. 3(b)].

\begin{figure}
\includegraphics[width=3.3in]{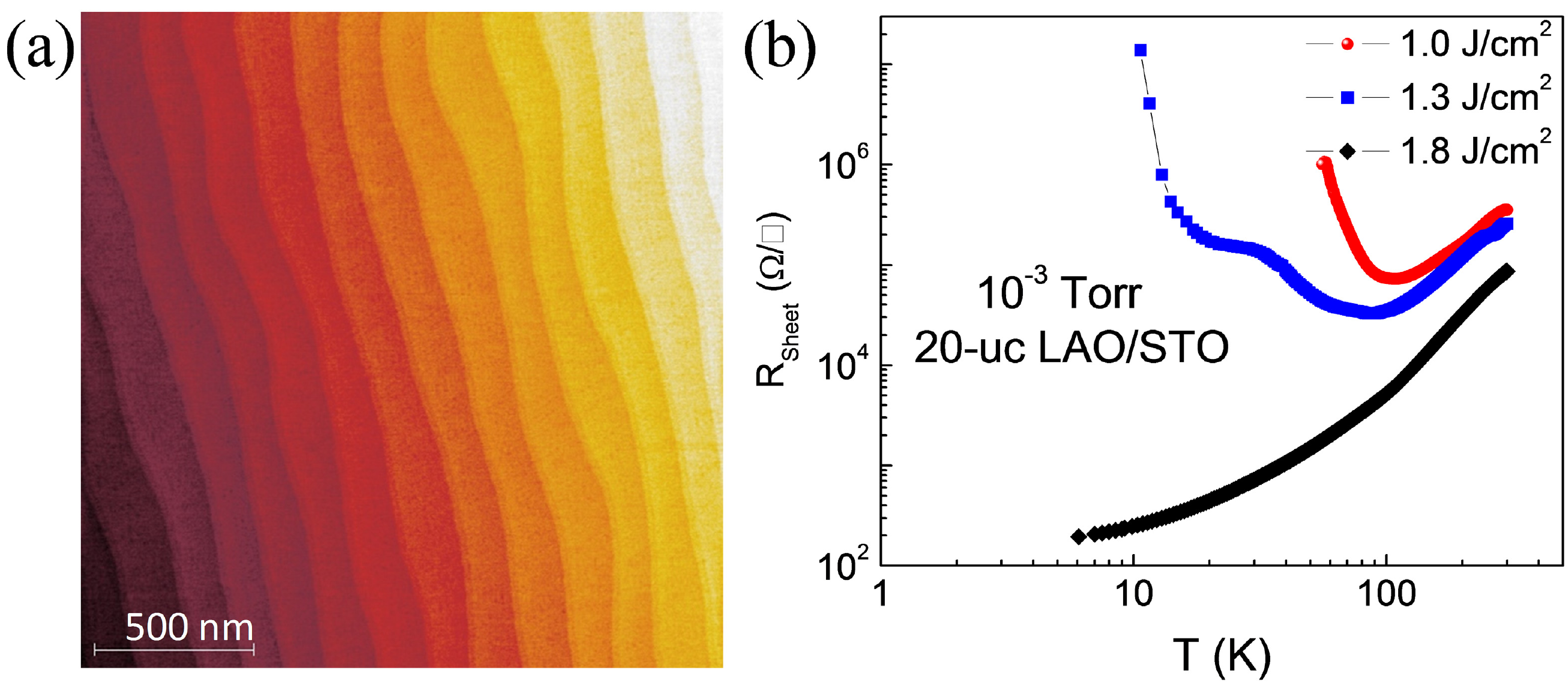}
\caption{\label{fig3}(Color online) (a) AFM image of an 20-uc LAO/STO heterostructure deposited on a TiO$_{2}$-terminaed STO substrate at 750 $^{\circ}$C in 10$^{-3}$ Torr oxygen pressure with the laser fluence of 1.3 J/cm$^{2}$. (b) Temperature-dependent sheet resistance of as-deposited 20-uc LAO/STO heterostructures fabricated with different laser fluence.}
\end{figure}

The recent study by Qiao \emph{et al.} revealed that the chemistry of LAO films directly affects the properties of LAO/STO heterostructures [15] and Breckenfeld \emph{et al.} [16] found that the composition of LAO films strongly depends on laser fluence. Low laser fluence results in Al-deficient LAO films, while high laser fluence leads to Al-rich LAO films. Al-rich films can remove more oxygen out of STO surface due to the strong chemical affinity of Al to oxygen atoms, thus creating more oxygen vacancies. That is why high-fluence-deposited 20-uc LAO/STO heterostructures are completely metallic in our case. Such result may not be able to differentiate whether the origin of low-temperature resistance minimum in crystalline LAO/STO heterostructures with thick LAO films is due to weak localization or the Kondo effect. However, it demonstrates that the low-temperature resistance minimum in LAO/STO heterostructure with thick LAO layers is largely tunable by the amount of oxygen vacancies in the system and can even be suppressed by increasing the amount of oxygen vacancies. These results provide new experimental perspective which could be useful in understanding the intrinsic mechanism of the low-temperature resistance minimum in the LAO/STO interface system.

In summary, we studied electrical properties of unannealed crystalline LAO/STO heterostructures fabricated at different conditions. It was found that the 2DEG-V formation in crystalline LAO/STO interface does not need 4-uc LAO and is independent of the polarization catastrophe. The crystalline LAO/STO heterostructures built on STO substrates with mixed terminations show metallic behavior for the 2DEG-V while they are insulating for the 2DEG-P. The low-temperature resistance minimum in crystalline LAO/STO heterostructures with thick LAO layers is largely tunable by laser fluence and thus the amount of oxygen vacancies in the system. All these results clearly demonstrate the dominant role of oxygen vacancies in electrical properties of unannealed crystalline LAO/STO heterostructures and fundamental differences between oxygen vacancies and polarization catastrophe in creating the 2DEG at crystalline LAO/STO interfaces.

\begin{acknowledgments}
We thank the National Research Foundation (NRF) Singapore under the Competitive Research Program (CRP) “Tailoring Oxide Electronics by Atomic Control” (Grant No. NRF2008NRF-CRP002-024), the National University of Singapore (NUS) for a cross-faculty grant, and FRC (ARF Grant No.R-144-000-278-112) for financial support.
\end{acknowledgments}


\begin{thebibliography}{99}

\bibitem{1}     A. Ohtomo and H. Y. Hwang, Nature {\bf 427},423 (2004).
\bibitem{2}     K. Nakagawa, H. Y. Hwang, and D. A. Muller, Nature Mater. {\bf 5}, 204 (2006).
\bibitem{3}     Z. Q. Liu, C. J. Li, W. M. L\"{u}, X. H. Huang, Z. Huang, S. W. Zeng, X. P. Qiu, L. S. Huang, A. Annadi, J. S. Chen, J. M. D. Coey, and T. Venkatesan, and Ariando, Phys. Rev. X {\bf 3}, 021010 (2013).
\bibitem{4}     Y. Chen, N. Pryds, J. E. Kleibeuker, G. Koster, J. Sun, E. Stamate, B. Shen, G. Rijnders, and S. Linderoth, Nano Lett. {\bf 11}, 3774 (2011).
\bibitem{5}     Z. Q. Liu, D. P. Leusink, X. Wang, W. M. L\"{u}, K. Gopinadhan, A. Annadi, Y. L. Zhao, X. H. Huang, S. W. Zeng, Z. Huang, A. Srivastava, S. Dhar, T. Venkatesan, and Ariando, Phys. Rev. Lett. {\bf 107}, 146802 (2011).
\bibitem{6}     S. Thiel, G. Hammerl, A. Schmehl, C. W. Schneider, and J. Mannhart, Science {\bf 313}, 1942 (2006).
\bibitem{7}     J. N. Eckstei, Nature Mater. {\bf 6}, 473 (2007).
\bibitem{8}     Ariando, X. Wang, G. Baskaran, Z. Q. Liu, J. Huijben, J. B. Yi, A. Annadi, A. R. Barman, A. Rusydi, S. Dhar, Y. P. Feng, J. Ding, H. Hilgenkamp, and T. Venkatesan, Nature Commun. {\bf 2}, 188 (2011).
\bibitem{9}     S. W. Lee, Y. Liu, J. Heo, and R. G. Gordon, Nano Lett. {\bf 12}, 4775 (2012).
\bibitem{10}    C. Cancellieri, N. Reyren, S. Gariglio, A. D. Caviglia, A. Fete, J.-M. Triscone, Europhys. Lett. {\bf 91}, 17004 (2010).
\bibitem{11}    J. Mannhart, D. H. A. Blank, H. Y. Hwang, A. J. Millis, and J.-M. Triscone, MRS Bulletin {\bf 33}, 1027 (2008).
\bibitem{12}    M. Huijben, Ph.D. Thesis, University of Twente, Enschede, The Netherlands, (2006).
\bibitem{13}    A. Brinkman, M. Huijben, M. van Zalk, J. Huijben, U. Zeitler, J. C. Maan, W. G. van der Wiel, G. Rijnders, D. H. A. Blank, and H. Hilgenkamp, Nat. Mater. {\bf 6}, 493 (2007).
\bibitem{14}    A. D. Caviglia, S. Gariglio, N. Reyren, D. Jaccard, T. Schneider, M. Gabay, S. Thiel, G. Hammerl, J. Mannhart, and J.-M. Triscone, Nature {\bf 456}, 624 (2008).
\bibitem{15}    L. Qiao, T. C. Droubay, T. Varga, M. E. Bowden, V. Shutthanandan, Z. Zhu, T. C. Kaspar, and S. A. Chambers, Phys. Rev. B {\bf 83}, 085408 (2011).
\bibitem{16}    E. Breckenfeld, N. Bronn, J. Karthik, A. R. Damodaran, S. Lee, N. Mason, and L. W. Martin, Phys. Rev. Lett. {\bf 110}, 196804 (2013).

\end{thebibliography}

\end{document}